\documentclass[twocolumn,aps,prl,groupedaddress,showpacs]
              {revtex4}%

\usepackage{graphicx}
\usepackage{amsmath}
\usepackage{bm}
\usepackage{relsize}
\RequirePackage{xspace}
\begin{document}


\title{\mbox{}\\[12pt]
Double charm production $\bm{e^+ e^- \to J/\psi + c \bar{c}}$ at B
factories with next-to-leading order QCD corrections}

\author{Yu-Jie Zhang$~^{(a)}$
  and Kuang-Ta Chao$~^{(a,b)}$}
\affiliation{ {\footnotesize (a)~Department of Physics, Peking
University,
 Beijing 100871, People's Republic of China}\\
{\footnotesize (b)~Center for High Energy Physics, Peking
University, Beijing 100871, People's Republic of China}}


\date{\today}

\begin{abstract}
The inclusive $J/\psi$ production in $e^+ e^- \to J/\psi c\bar c$ at
B factories is one of the most challenging open problems in heavy
quarkonium physics. The observed cross section of this double charm
production process is larger than existing leading order (LO) QCD
predictions by a factor of 5. In the nonrelativistic QCD (NRQCD)
factorization formalism, we calculate the next-to-leading order
(NLO) QCD virtual and real corrections to this process, and find
that these corrections can substantially enhance the cross section
with a $K$ factor of about 1.8. We further take into account the
feeddown contributions from higher charmonium states (mainly the
$\psi(2S)$ as well as $\chi_{cJ}$) and the two-photon contributions,
and find that the discrepancy between theory and experiment can be
largely removed.

\end{abstract}

\pacs{13.66.Bc, 12.38.Bx, 14.40.Gx}

\maketitle


The double charm production in $e^+e^-$ annihilation at B factories
is one of the most challenging open problems in heavy quarkonium
physics and nonrelativistic QCD (NRQCD) (for a review of related
problems, see\cite{Brambilla:2004wf}). The exclusive production
cross section of double charmonium in $e^+ e^- \to J/\psi \eta_c$ at
$\sqrt{s}=10.6~$GeV measured by
%
Belle \cite{Abe:2002rb, Pakhlov:2004au}
 and BaBar\cite{BaBar:2005}
are larger than the leading order (LO) calculations $3.8\sim
5.5$fb~\cite{Braaten:2002fi,Liu:2002wq} in NRQCD by possibly almost
an order of magnitude (see also \cite{hagiwara}).

Moreover, the inclusive  $J/\psi$ production cross section via
double $c\bar c$ in $e^+ e^- \to J/\psi c\bar c$ at
$\sqrt{s}=10.6~$GeV measured by Belle\cite{Abe:2002rb}
\begin{eqnarray}\label{bellecc}
\sigma[e^+e^-\to J/\psi + c \bar c+X] = \left(
0.87^{+0.21}_{-0.19}\pm 0.17 \right) \; {\rm pb}~,
\end{eqnarray}
is about a factor of 5 higher than theoretical predictions including
both the color-singlet\cite{cs,Liu:2003jj} and
color-octet\cite{Liu:2003jj} $c\bar c$ contributions at leading
order(LO) of $\alpha_s$ in the NRQCD factorization
formalism\cite{BBL}.
This is another intriguing challenge in the double charm
production problem, aside from the exclusive $J/\psi\eta_c$
production in $e^+e^-$ annihilation.

Some theoretical studies were attempted in order to resolve the
large discrepancy in $e^+ +e^- \to J/\psi +c \bar c$. Liu, He, and
Chao calculated the color-octet contribution~\cite{Liu:2003jj} and
the two-photon contribution to $J/\psi +c \bar c$
production~\cite{Liu:2003zr} in NRQCD. But those contributions are
small and can not make up such a large discrepancy. Hagiwara et al.
assumed a large renormalization K factor $(K\sim 4)$ for the $J/\psi
c \bar c$ cross section\cite{Hagiwara:2004pf}. Kaidalov introduced a
nonperturbative quark-gluon-string model\cite{Kaidalov:2003wp}. Kang
et al. got
 $\sigma(e^+e^-\to J/\psi +c\bar{c}+X)/\sigma(e^+e^-\to J/\psi +X)=%
0.049$ in the color-evaporation-model \cite{Kang:2004zj}.  Other
suggestions to resolve this problem may be found in
Ref.~\cite{Brambilla:2004wf}.

In order to further clarify this problem, in this paper we present a
result for  the next-to-leading order (NLO) QCD correction to the
inclusive $J/\psi$ production process of $e^+ + e^- \to J/\psi+c
\bar c$. And we have already found that the NLO QCD correction to
the exclusive production process of $e^+ + e^- \to J/\psi+\eta_c$ is
crucial, for which the K factor (the ratio of NLO to LO ) may reach
to a value of about 2, and hence essentially reduces the large
discrepancy between theory and experiment of $e^+ + e^- \to
J/\psi+\eta_c$\cite{Zhang:2005ch} (with significant relativistic
corrections further considered this discrepancy is probably
resolved\cite{bodwin,he}; enhancement effects due to use of
light-cone formalism and relativistic corrections are also proposed
in \cite{ma}).

\begin{figure}
\includegraphics[width=8.5cm]{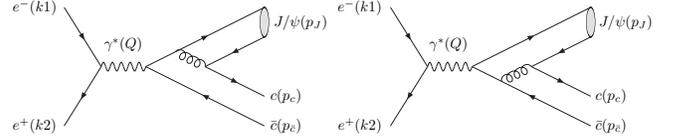}
\caption{\label{fig1}Two of the four Born diagrams for $e^- e^+ \to
J/\psi c \bar c$. }
\end{figure}

At LO in $\alpha_s$, $J/\psi + c \bar c$ can be produced at order
$\alpha^2\alpha_s^2$ (see, e.g.,
Refs.~\cite{Liu:2002wq,Zhang:2005ch}). There are four Feynman
diagrams, two of which are shown in Fig.~\ref{fig1}, and the other
two diagrams can be found through reversing the arrows on the quark
lines. Momenta for the involved particles are assigned as $e^-(k_1)
e^+ (k_2)\to \gamma^*(Q) \to J/\psi (p_J)+ c(p_c)+ \bar c(p_{\bar
c})$. In the calculation, we use {\tt FeynArts}~\cite{feynarts} to
generate Feynman diagrams and amplitudes; {\tt
FeynCalc}~\cite{Mertig:an} for the tensor reduction; and {\tt
LoopTools}~\cite{looptools} for the numerical evaluation of the
infrared (IR)-safe one-loop integrals. Finally we use {\tt
Mathematica} to integrate over phase space and get the numerical
results.

To NLO calculation, the cross section is
\begin{eqnarray}
\sigma &=&\sigma_{Born} + \sigma_{virtual} + \sigma_{real}+
{\cal{O}}(\alpha^2 \alpha_s^4),
\end{eqnarray}
where
\begin{eqnarray}\label{crosssections}
{\rm d}\sigma_{Born} &=& \frac{1}{4}\ \frac{1}{2s}\ \sum
|\mathcal{M}_{Born} |^2
\rm{d}PS_3,\nonumber \\
{\rm d}\sigma_{virtual} &=&\frac{1}{4}\ \frac{1}{2s} \ \sum \ 2 \
{\rm
Re}(\mathcal{M}_{Born}^*\mathcal{M}_{NLO}) \rm{d}PS_3,\nonumber \\
{\rm d}\sigma_{real} &=&\frac{1}{4}\ \frac{1}{2s}\ \sum
|\mathcal{M}_{real} |^2 \rm{d}PS_4.
\end{eqnarray}
Here the factor $1/2s$ is the flux factor. $\sum$ means sum over the
polarizations of initial and final state particles. $\rm{d}PS_3$ and
$\rm{d}PS_4$ are the three and four-body phase spaces respectively.

\begin{figure*}
\includegraphics[width=17.5cm]{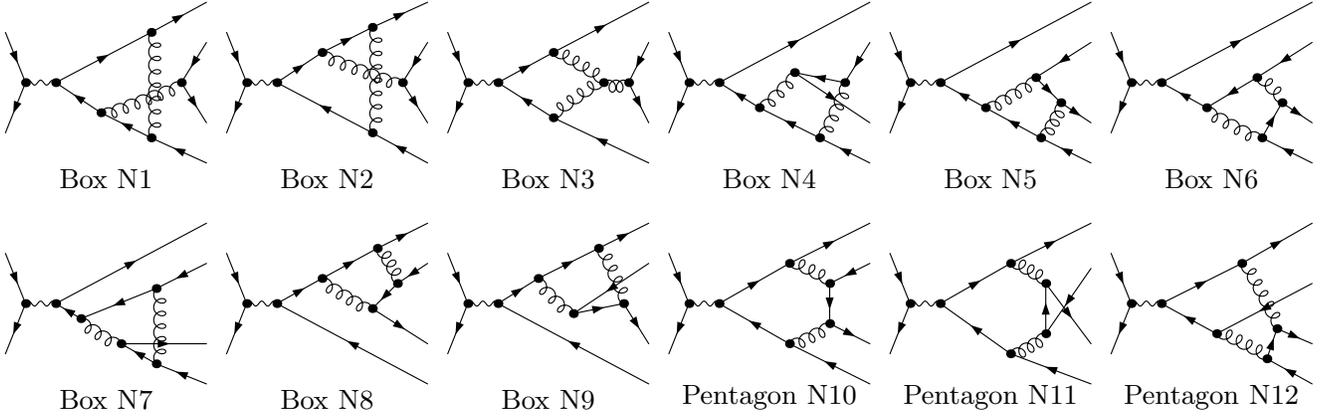}
\caption{\label{fig2} Twelve of the twenty-four box and pentagon
diagrams for $e^-(k_1) e^+ (k_2)\to J/\psi(p_J)+ c(p_c) \bar
c(p_{\bar c})$. }
\end{figure*}

There are ultraviolet(UV), infrared(IR), and Coulomb singularities,
and we treat them in the same way as in Ref.~\cite{Zhang:2005ch}.
For the box diagrams shown in Fig.~\ref{fig2}, $\mathrm{Box N8}$ and
$\mathrm{ Box N10 } $ have IR and Coulomb singularities,
$\mathrm{Box N3 } $ does not have IR singularity, while the other
nine diagrams have IR singularity. $\mathrm{Box N1 +Box N4 } $ ,
$\mathrm{Box N6 +Box  N7 +Pentagon N12} $ are IR finite
respectively. IR terms of $\mathrm{ Box N9+Box N2 +Pentagon N11}$
are canceled by Vertex diagrams. The IR term in counter terms and
$\mathrm{ Box N5+ Box N8 +Pentagon N10} $ should be canceled by the
real corrections. And the Coulomb singularity terms in $\mathrm{ Box
N8 +Pentagon N10} $ should be mapped into the wave functions of
$J/\psi$. Similar to Ref.~\cite{Zhang:2005ch}, we use
$D=4-2\epsilon$ space-time dimension and the relative velocity $v$
to regularize the IR and Coulomb singularities. The IR and Coulomb
singularity terms in the virtual corrections are
 \begin{eqnarray}
&&d\sigma_{virtual} ^{IR,Coulomb}\nonumber \\ &=&
d\sigma_{Born}\frac{ 4 \alpha_s}{3\pi} \left(
\frac{\pi^2}{v}-\frac{1}{ \epsilon}- \frac{ p_c \cdot p_{\bar c}
\ln(-\bar s_{c \bar c})}{\sqrt{(p_c \cdot p_{\bar c})^2-m^4}}
\frac{1}{\epsilon} \right),\end{eqnarray} where $\bar s_{c \bar c} =
\frac{\sqrt{1-4m^2/(p_c +p_{\bar c})^2 }-1} {\sqrt{1-4m^2/(p_c
+p_{\bar c})^2 }+1}$.

\begin{figure*}
\includegraphics[width=17.5cm]{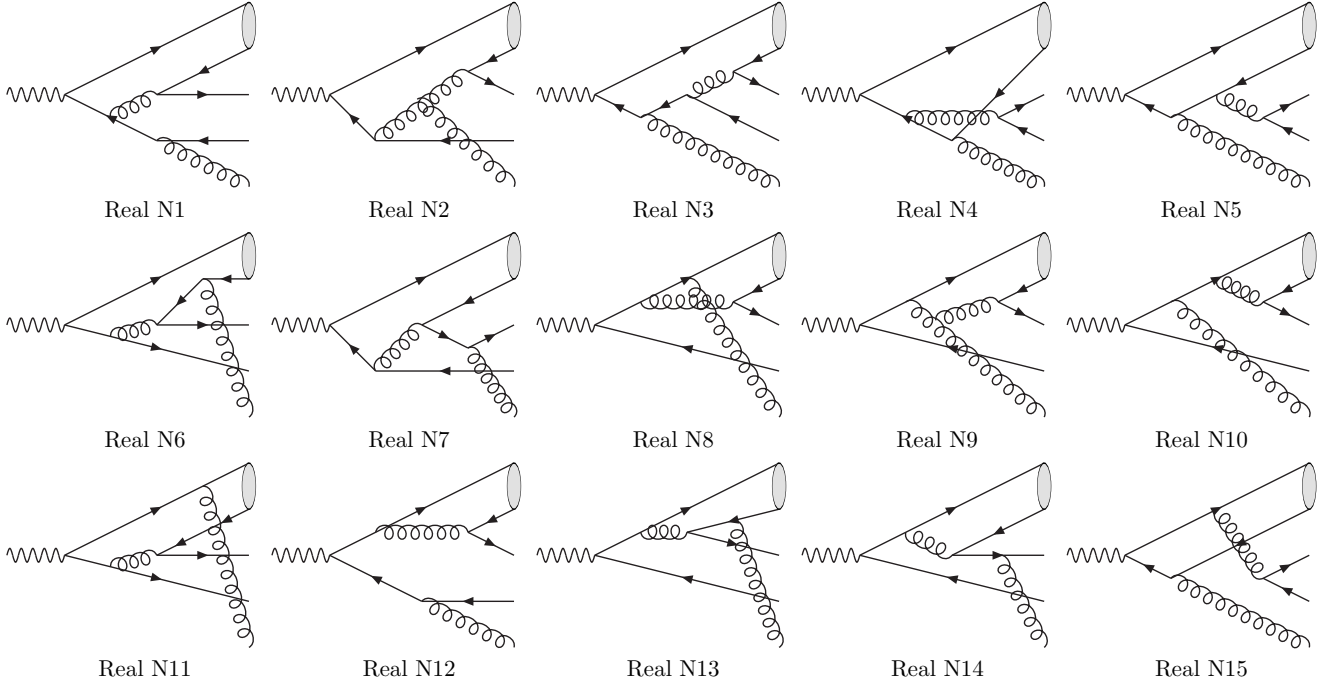}
\caption{\label{real} Fifteen of the thirty real correction diagrams
for $e^- e^+ \to J/\psi+ c \bar c+g$. }
\end{figure*}
There are thirty diagrams for real corrections, and half of them are
shown in Fig.\ref{real}. The other fifteen diagrams can be obtained
through reversing the arrows on the quark lines that are connected
with $J/\psi$. The calculation of the real corrections is similar to
the leading order calculation, but there should appear the IR
singularity \cite{Harris:2001sx}.
We find that ${\rm Real\ N1}$, ${\rm Real \ N7 }$, ${\rm Real \ N12
}$, and ${\rm Real\ N14 }$ are associated with IR singularity. And
the eikonal factors of ${\rm Real\ N4}$, ${\rm Real\ N6}$, ${\rm
Real\ N8}$, ${\rm Real\ N11}$, ${\rm Real\ N13}$, and ${\rm Real\
N15}$, in which the gluon is connected with the external charm quark
and anti-charm quark in the $J/\psi$, are canceled by themselves.
The other five diagrams ${\rm Real\ N2}$, ${\rm Real\ N3}$, ${\rm
Real\ N5}$, ${\rm Real\ N9}$, and ${\rm Real\ N10}$ are independent
of IR singularity. Using the eikonal approximation, we get the IR
singularity terms in real corrections
\begin{eqnarray}
d\sigma_{real}^{IR}=d\sigma_{Born} \frac{4\alpha_s}{3
\pi}\frac{1}{\epsilon} \left(1+ \frac{ \ln(-\bar s_{c \bar c}) p_c
\cdot p_{\bar c}}{\sqrt{(p_c \cdot p_{\bar c})^2 -m^4}}\right)
\end{eqnarray}
They just cancel the IR singular terms of
$d\sigma_{virtual}^{IR,Coulomb}$. The Coulomb singular terms in
$d\sigma_{virtual}^{IR,Coulomb}$ can be mapped into the $J/\psi$
wave function.

We now turn into numerical calculations for the cross section of
$e^+ + e^-\rightarrow J/\psi+c \bar c+X$. To be consistent with the
above NLO calculation, the value of the $J/\psi$ wave function
squared at the origin should be extracted from the leptonic width at
NLO of $\alpha_s$ (see e.g.~\cite{BBL})
\begin{equation}
|R_S(0)|^2=\frac{9m^2_{J/\psi}}{16\alpha^2(1-4 C_F \alpha_s/
\pi)}\Gamma(J/\psi\to e^+e^-). \label{eq:14}
\end{equation}
Using the experimental value $5.55\pm 0.14\pm 0.02$~KeV~\cite{PDG},
we obtain $|R_S(0)|^2=1.01 \mbox{GeV}^3$, which is a factor of 1.25
larger than $0.810 \mbox{GeV}^3$ that was used in
Ref.\cite{Liu:2003jj} from potential model calculations. Taking
$m_{J/\psi}=2m$ (in the nonrelativistic limit),
 $m\!=\!1.5~$GeV, $\Lambda^{(4)}_{\overline{\rm MS}}=338{\rm MeV}$,
  $\alpha_s(\mu)=0.259$ for $\mu=2m$ (these
are the same as in Ref.\cite{Liu:2003jj} except here a larger
$|R_S(0)|^2$ is used), the cross section for $e^+ + e^-\rightarrow
J/\psi+c \bar c +X$ at NLO of $\alpha_s$ is
\begin{equation}
\label{jsetac} \sigma(e^+ + e^-\rightarrow J/\psi+c \bar c+X )=0.33\
\rm{pb}.
\end{equation}
It is a factor of $1.8$ larger than the LO result $0.18$~pb obtained
with the same parameters. For $\mu=m$ and $\sqrt{s}/2$, we have
$\alpha_s=0.369$ and $0.211$, and get the cross section $0.53$~pb
and $0.24$~pb respectively.
\begin{figure}
\includegraphics[width=8.5cm]{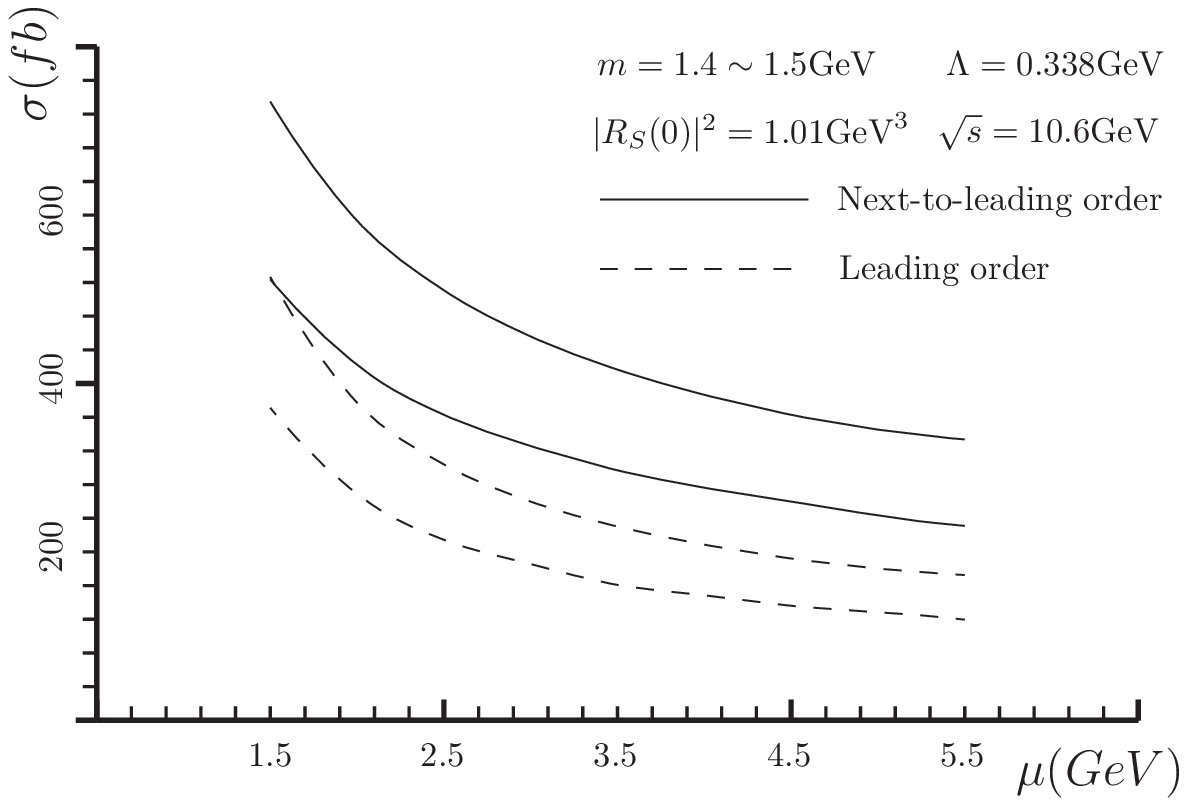}
\caption{\label{dependenceOFmu}Direct production cross sections of
$e^+ + e^-\rightarrow J/\psi+c \bar c+X$ as functions of the
renormalization scale $\mu$. Here $|R_S(0)|^2=1.01 {\rm GeV}^3$,
$\Lambda=0.338 {\rm GeV}$, $ \sqrt s=10.6{\rm GeV} $; NLO results
are represented by solid lines and LO one by dashed lines; the upper
line is for $m=1.4{\rm GeV}$ and the corresponding lower line is for
$ m=1.5{\rm GeV} $.}
\end{figure}
 If we set $m=1.4{\rm
GeV}$ and $\mu=2m$, the cross section at NLO of $\alpha_s$ is
\begin{equation}
\label{jsccc} \sigma(e^+ + e^-\rightarrow J/\psi+c \bar c+X )=0.47\
\rm{pb}.
\end{equation}
It is about a factor of $1.7$ larger than leading order cross
section $0.27$~pb.  The dependence of the cross section on the
renormalization scale $\mu$ is shown in Fig.~\ref{dependenceOFmu}.
When $\mu$ changes from $m_c=1.5~$GeV to $\sqrt s/2=5.3~$GeV, the
ratio $\sigma(\mu)/\sigma(\sqrt s/2)$ is found to vary from $3.05$
to $1$ in LO result. For NLO, with $m_c=1.4(1.6)~$GeV, the ratio
$\sigma(\mu)/\sigma(\sqrt s/2)$ varies from $2.22 (2.26)$ to $1$. We
see that, as expected, the scale-dependence in NLO is considerably
reduced compared with that in the LO. A detailed discussion
will be given elsewhere.


We should also include the QED contribution as well as the
two-photon contribution of $e^+e^- \to 2\gamma^* \to \ J/\psi+ c
{\bar c}$. Furthermore, since the experimental data are for the
prompt $J/\psi+ c {\bar c}+X$ production, we should consider the
feeddown contributions from higher charmonium states such as $e^+e^-
\to\psi(2S)+ c {\bar c} +X\to J/\psi+ c {\bar c}+X$, and $e^+e^-
\to\chi_{cJ}+ c {\bar c} \to J/\psi+ c {\bar c}+X$.

\begin{figure}
\includegraphics[width=8.5cm]{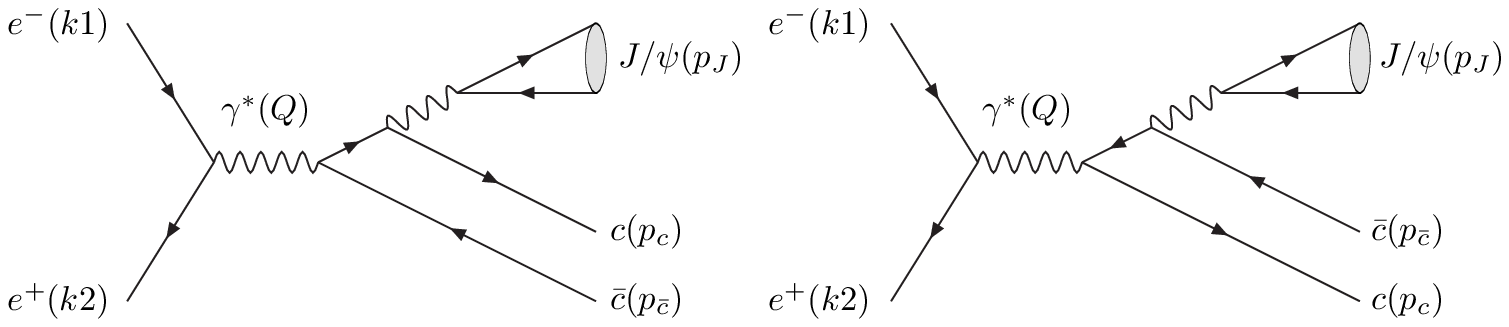}
\caption{\label{JpsccbarQED}Two of the six QED diagrams for $e^+e^-
\to J/\psi c \bar c$. }
\end{figure}

Two of the six QED diagrams of $e^+e^- \to \gamma^* \to \ J/\psi+ c
{\bar c}$ are shown in Fig.~\ref{JpsccbarQED}. The other four
diagrams can be obtained by replacing the gluon with the photon
shown in Fig.~\ref{fig1}. Contributions from QED diagrams can
interfere with that from QCD Born diagrams, and resulted in a cross
section of $8~ {\rm fb}$ at order ${\cal O}(\alpha_s \alpha^3)$.

$e^+e^- \to 2\gamma^* \to J/\psi c {\bar c}$ has been calculated by
Liu et al.\cite{Liu:2003zr}. Using their result, this cross section
is $23 \times \frac{1.01}{0.810}\ {\rm fb}=29\ {\rm fb}$, where
the factor $\frac{1.01}{0.810}$ is due to using the new value of the
$J/\psi$ wave function at the origin.

At leading order in $v$ (the relative velocity of quark and
antiquark in the charmonium rest frame), the difference between
$e^+e^- \to\psi(2S)+ c {\bar c}$ and $e^+e^- \to J\psi+ c {\bar c}$
is in the wave functions at the origin. Using Eq.(\ref{eq:14}), the
contribution from the transition $\psi(2S) \to J/\psi$ is to enlarge
the direct production of $J/\psi+ c {\bar c}+X$ by a factor of $
\frac{|R_{2S}(0)|^2}{|R_{1S}(0)|^2} \mathrm{B}(\psi(2S) \to J/\psi
X). $ By using $\Gamma(\psi(2S)\to e^+e^-)= 2.48\pm 0.06 {\rm KeV}$
and the branching ratio for the $\psi(2S)\to J/\psi X$ transition
fraction $B=56.1\pm 0.9\%$~\cite{PDG}, we find the enlarging factor
to be $0.355 $.

The cross sections of $e^+e^- \to \chi_{cJ}+ c {\bar c}+X$ with both
color singlet and octet contributions were calculated in
Ref.\cite{Liu:2003jj}. Moreover, the color octet contribution to
$J/\psi c\bar c$ production was also estimated to be about $11\
\rm{fb}$ in Ref.\cite{Liu:2003jj}. Using their results and the
observed branching ratios for $\chi_{cJ} \to J/\psi \gamma$
transitions $B=1.31\%,35.6\%,20.2\%$ for $J=0,1,2$
respectively~\cite{PDG}, we find the sum of feeddown from
$\chi_{cJ}$ and color-octet for $J/\psi$ contributions to be $21\
\rm{fb}$.

Combining all these contributions, the prompt cross section of
$e^+e^- \to J/\psi+ c {\bar c}+X$ at NLO of $\alpha_s$ is
\begin{equation}
\label{jsetacProm} \sigma_{prompt}(e^+ + e^-\rightarrow J/\psi+c
\bar c+X )=0.51\ \rm{pb}.
\end{equation}
It is $59\%$ of the data value in Eq.~(\ref{bellecc}). If we set
$m=1.4{\rm GeV}$ and $\mu=2m$, and ignore the differences of other
contributions due to the change of mass, then the prompt cross
section of $e^+e^- \to J/\psi+ c {\bar c}+X$ at NLO of $\alpha_s$ is
\begin{equation}
\label{jsetac} \sigma_{prompt}(e^+ + e^-\rightarrow J/\psi+c \bar
c+X )=0.71\ \rm{pb}.
\end{equation}
It is $82\%$ of the experimental value in Eq.~(\ref{bellecc}).

In order to see the uncertainties due to parameters
$m_c,~\mu,~\alpha_s(\mu)$, we set $\mu=2.8(5.3)~$GeV,
$m=1.4\mp0.2~$GeV, and then get the cross section in the range
$0.71^{+0.94}_{-0.31}(0.53^{+0.59}_{-0.23})~$pb.
%
The NLO QCD correction to $e^+ + e^-\rightarrow J/\psi+c \bar c$ is
large, despite of existing parametric uncertainties.

%
%

In conclusion, we find that by taking into consideration all NLO
virtual corrections with self energy, triangle, box, and pentagon
diagrams, and the real corrections, and factoring the Coulomb
singular term into the $c\bar c$ bound state wave function, we get
an ultraviolet (UV) and infrared (IR) finite correction to the
direct production cross section of $e^+e^-\to J/\psi + c \bar c$ at
$\sqrt{s}=10.6$~GeV, and find that the NLO QCD correction can
substantially enhance the cross section with a K factor of about
1.8. With $m=1.4{\rm GeV}$ and $\mu=2m$, the cross section of direct
$J/\psi c\bar c$ production through one-photon is estimated to be
$0.47$~pb. Adding the feeddown contributions from higher charmonium
states (mainly the $\psi(2S)$ as well as $\chi_{cJ}$) and
contributions from two-photon process and color-octet channels, the
prompt production cross section of $e^+e^-\to J/\psi + c \bar c$ at
NLO in $\alpha_s$ is found to be $0.71$~pb, which is $82\%$ of the
experimental value $0.87$~pb. Hence the discrepancy between theory
and experiment is largely removed, despite of certain theoretical
uncertainties.

\begin{acknowledgments}
This work was supported in part by the National Natural Science
Foundation of China (No 10421503, No 10675003), the Key Grant
Project of Chinese Ministry of Education (No 305001), and the
Research Found for Doctorial Program of Higher Education of China.
\end{acknowledgments}



\begin{thebibliography}{}

\bibitem{Brambilla:2004wf}
N.~Brambilla {\it et al.}, Yellow Report on Heavy Quarkonium
Physics, arXiv:hep-ph/0412158.


\bibitem{Abe:2002rb}
K.~Abe {\it et al.} [BELLE Collaboration],
Phys.\ Rev.\ Lett.\ {\bf 89}, 142001 (2002).




\bibitem{Pakhlov:2004au}
  P.~Pakhlov  [Belle Collaboration],
  arXiv:hep-ex/0412041.






\bibitem{BaBar:2005}
  B.~Aubert {\it et al.}  [BABAR Collaboration],
  Phys.\ Rev.\ D {\bf 72}, 031101 (2005)






\bibitem{Braaten:2002fi}
E.~Braaten and J.~Lee,
Phys.\ Rev.\ D {\bf 67}, 054007 (2003)

\bibitem{Liu:2002wq}
K.~Y.~Liu, Z.~G.~He and K.~T.~Chao,
Phys.\ Lett.\ B {\bf 557}, 45 (2003);
~hep-ph/0408141.

\bibitem{hagiwara}
K. Hagiwara, E. Kou, and C.F. Qiao, Phys. Lett. B570, 39 (2003).


\bibitem{cs}
P.~Cho and A.~K.~Leibovich,
  Phys.\ Rev.\ D {\bf 54}, 6690 (1996)
  F.~Yuan, C.~F.~Qiao and K.~T.~Chao,
  Phys.\ Rev.\ D {\bf 56}, 321 (1997)
  F.~Yuan, C.~F.~Qiao and K.~T.~Chao,
  Phys.\ Rev.\ D {\bf 56}, 1663 (1997)
  S.~Baek, P.~Ko, J.~Lee and H.~S.~Song,
  J.\ Korean Phys.\ Soc.\  {\bf 33}, 97 (1998)
  V.~V.~Kiselev, A.~K.~Likhoded and M.~V.~Shevlyagin,
  Phys.\ Lett.\ B {\bf 332}, 411 (1994)




\bibitem{Liu:2003jj}
  K.~Y.~Liu, Z.~G.~He and K.~T.~Chao,
  Phys.\ Rev.\ D {\bf 69}, 094027 (2004)


\bibitem{BBL}
G.~T.~Bodwin, E.~Braaten, and G.~P.~Lepage,
    Phys.\ Rev.\ D {\bf 51}, 1125 (1995);
{\bf 55}, 5853(E) (1997).






\bibitem{Liu:2003zr}
  K.~Y.~Liu, Z.~G.~He and K.~T.~Chao,
  Phys.\ Rev.\ D {\bf 68}, 031501(R) (2003)


\bibitem{Hagiwara:2004pf}
  K.~Hagiwara, E.~Kou, Z.~H.~Lin, C.~F.~Qiao and G.~H.~Zhu,
  Phys.\ Rev.\ D {\bf 70}, 034013 (2004)




\bibitem{Kaidalov:2003wp}
  A.~B.~Kaidalov,
  JETP Lett.\  {\bf 77}, 349 (2003)



\bibitem{Kang:2004zj}
  D.~Kang, J.~W.~Lee, J.~Lee, T.~Kim and P.~Ko,
  Phys.\ Rev.\ D {\bf 71}, 094019 (2005)






\bibitem{Zhang:2005ch}
  Y.~J.~Zhang, Y.~J.~Gao and K.~T.~Chao,
  Phys.\ Rev.\ Lett.\  {\bf 96}, 092001 (2006)

\bibitem{bodwin}
G.T. Bodwin, D. Kang, and J. Lee, Phys. Rev. D74, 014014 (2006);
G.T. Bodwin et al., hep-ph/0611002.

\bibitem{he}
Z.G. He, Y. Fan, and K.T. Chao, hep-ph/0702239.

\bibitem{ma}
J.P. Ma and Z.G. Si, Phys. Rev. D70, 074007 (2004); hep-ph/0608221;
V.V. Braguta, A.K. Likhoded, and A.V. Luchinsky, Phys. Rev. D72,
074019 (2005); hep-ph/0602047; hep-ph/0611021; A.E. Bondar and V.L.
Chernyak, Phys. Lett. B612, 215 (2005); D. Ebert and A.P.
Martynenko, Phys. Rev. D74, 054008 (2006).

\bibitem{feynarts} M. B\"ohm, A. Denner, J. K\"ublbeck,
Comput.\ Phys.\ Commun.\ {\bf 60 }(1990) 165;
  T.~Hahn,
  Comput.\ Phys.\ Commun.\  {\bf 140}, 418 (2001)




\bibitem{Mertig:an} R. Mertig, M. B\"ohm, A. Denner,
Comput.\ Phys.\ Commun.\  {\bf64 }(1991) 345.


\bibitem{looptools}
  T.~Hahn and M.~Perez-Victoria,
  Comput.\ Phys.\ Commun.\  {\bf 118}, 153 (1999)



\bibitem{Harris:2001sx}
  B.~W.~Harris and J.~F.~Owens,
  Phys.\ Rev.\ D {\bf 65}, 094032 (2002)


\bibitem{PDG}
  W.~M.~Yao {\it et al.}  [Particle Data Group],
  J.\ Phys.\ G {\bf 33}, 1 (2006).


\end{thebibliography}
\end{document}